\documentstyle[12pt,aasms4]{article}

\lefthead{Yasuhiro Shioya \& Kenji Bekki}
\righthead{SED evolution of dusty starburst}

\begin{document}
\title{Effects of dust extinction on optical spectroscopic 
properties for starburst galaxies in distant clusters}

\author{Yasuhiro Shioya} 
\affil{Astronomical Institute, Tohoku University, Sendai, 980-8578, Japan} 

\and

\author{Kenji Bekki}
\affil{Division of Theoretical Astrophysics, National Astronomical 
Observatory, Mitaka, Tokyo, 181-8588, Japan}

\begin{abstract}
Recent observational studies on galaxies in distant clusters 
discovered a significant fraction of possible dusty starburst 
galaxies with the so-called `e(a)' spectra that are characterized 
by strong H$\delta$ absorption and relatively modest [OII] emission. 
We numerically investigate spectroscopic and photometric 
evolution of dusty starburst galaxies in order to clarify the 
origin of the e(a) spectra. 
We found that if a young starburst population is 
preferentially obscured by dust than an old one in 
a dusty starburst galaxy, the galaxy shows e(a) spectrum. 
It is therefore confirmed that 
the selective dust extinction, which is first suggested by 
Poggianti \& Wu (2000) and means the strongest dust extinction for 
the youngest stellar population among stellar populations with 
different ages, is critically important to 
reproduce quantitatively the observed e(a) spectra 
for the first time in the present numerical study. 
We furthermore discuss what 
physical process is closely associated with this selective 
dust extinction in cluster environment. 
\end{abstract}


\section{Introduction}

Since Butcher \& Oemler (1978) discovered a large fraction of blue 
galaxies in distant clusters of galaxies at the redshift $z \ge 0.2$, 
the origin of the blue color and physical processes closely associated 
with the formation of the distant blue populations have been 
extensively discussed by many authors 
(Dressler \& Gunn 1983, 1992; Lavery \& Henry 1986; 
Couch et al. 1994, 1998; Abraham et al. 1996; Barger et al. 1996; 
Fisher et al. 1998; Morris et al. 1998; Balogh et al. 1999). 
In particular, Dressler \& Gunn (1983, 1992) discovered galaxies 
with no detectable emission lines and very strong Balmer absorption ones 
(the so-called E+A galaxies) and concluded that some fraction of 
distant blue populations are poststarburst galaxies that 
truncated abruptly their active star formation. 
One of longstanding and remarkable problems concerning 
the evolution of galaxies 
in cluster environment is to understand clearly a physical relationship 
between these blue populations,
whose number fractions rapidly increase 
with redshift, and passive populations such as elliptical galaxies 
and S0s. 
Several attempts have been made to clarify an evolutionary link 
among star-forming, poststarburst, and passively evolving galaxies 
observed in distant clusters of galaxies (Couch \& Sharples 1987; 
Abraham et al. 1996; Barger et al. 1996; Morris et al. 1998; 
Balogh et al. 1999). 
It is however not so clear whether there is really an evolutionary link 
among these populations with various spectroscopic 
properties and what physical process can drive the evolution. 

Recent observational studies have found a significant population of possible 
dusty starburst galaxies in distant clusters 
(Poggianti et al. 1999: Owen et al. 1999; 
Smail et al. 1999), which provide a new clue to 
the evolutionary link among various spectral classes. 
Poggianti et al. (1999) suggested that galaxies with strong Balmer 
absorption and relatively modest [OII] emission, which are classified as 
``e(a)'' galaxies by Dressler et al. (1999), are dusty starburst galaxies 
and furthermore discussed that distant clusters have a significant fraction 
of these dusty starburst galaxies. 
Smail et al. (1999) found that spectral properties of 5 out of 10 galaxies 
detected by 1.4 GHz VLA radio observation are 
classified as poststarburst (a+k/k+a type in Dressler et al. 1999) 
and considered that the star formation is hidden by dust in them.  
This finding suggested that dust effects are remarkable even for galaxies that 
were previously identified as poststarburst ones by optical spectral 
properties. 
Although a growing number of observational results on e(a) galaxies with 
possible dusty starburst have been accumulated, there are only a few 
theoretical studies addressing the formation and evolution of 
these e(a) populations in distant clusters. 

The purpose of this Letter is to investigate the origin and the nature 
of e(a) population observed in distant clusters of galaxies by using a 
one-zone chemical and spectrophotometric evolution model. 
We particularly investigate whether dusty starburst galaxies can show 
both strong H$\delta$ absorption and relatively modest [OII] emission 
observed in distant e(a) galaxies. 
We demonstrate that if a younger stellar population formed  
during a starburst is preferentially obscured by dust  
in a galaxy, spectral properties of the galaxy during  
starburst become very similar to those characteristics of e(a) 
populations. 
We refer to this behavior of dust 
extinction as {\it a selective dust extinction}, which is  
originally proposed by Poggianti and Wu (2000) and means that 
the effect of dust extinction is maximum for the youngest stellar 
populations among stellar populations with different ages. 
We furthermore suggest that this selective extinction can be 
achieved in interacting and merging galaxies in distant clusters. 
In the followings, the cosmological parameters $H_0$ and $q_0$ are set to be 
$65 {\rm km \; s^{-1} \; Mpc^{-1}}$ and 0.05 respectively, which 
means that the corresponding present age of the universe is 13.8 Gyr. 

\section{Model}

We adopt a one-zone chemical and spectrophotometric evolution model 
of a disk galaxy with a starburst, and thereby investigate 
when and how a disk galaxy shows strong H$\delta$ absorption line and 
relatively modest [OII] emission one characteristics of e(a) spectra 
observed in distant clusters of galaxies. 
Since more details of the adopted model are given in Shioya \& Bekki 
(1998, 2000 in preparation), 
we only briefly describe the model in the present study. 
We follow the chemical evolution of galaxies by using the model 
described in Matteucci \& Tornamb\`{e} (1987) which includes 
metal-enrichment processes of type Ia and II supernovae (SNIa and SNII). 
We adopt the Salpeter initial mass function (IMF), 
$\phi(m) \propto m^{-1.35}$, with upper mass limit 
$M_{\rm up}=120M_{\odot}$ and lower mass limit 
$M_{\rm low}=0.1M_{\odot}$. 
We calculate photometric properties of galaxies as follows. 
The monochromatic flux of a galaxy with age $T$, $F_{\lambda}(T)$, 
is described as 
\begin{equation}
F_{\lambda} (T) = \int_0^T F_{\rm SSP,\lambda}(Z,T-t) \psi(t) dt, 
\end{equation}
where $F_{\rm SSP,\lambda}(Z,T-t)$ is a monochromatic flux of 
a single stellar population with age $T-t$ and metallicity $Z$, and 
$\psi(t)$ is time-dependent star formation rate described later. 
In the present study, we use the spectral library GISSEL96 which is the 
latest version of Bruzual \& Charlot (1993). 

The star formation history of a disk galaxy is characterized by three epochs.  
The first is the epoch of galaxy formation 
($z_{\rm form}$ in redshift and the age of galaxy is 0 Gyr) 
at which a disk galaxy forms and begins 
to consume initial interstellar gas by star formation with the moderate rate. 
In the present study, $z_{\rm form}$ is fixed at 4.5 and  
therefore the age of galaxies at $z=0$ is 12 Gyr. 
The second is $z_{\rm sb}$ at which starburst begins in the disk.  
In the present study, $z_{\rm sb}$ is set at 0.4 and the age of galaxies 
at $z_{\rm sb}$, $T_{\rm sb}$, is 7.64 Gyr. 
The third is $z_{\rm end}$ at which star formation ceases and   
is defined as the epoch at which stellar mass fraction 
becomes 0.95 in our models. 
The age of galaxies at $z_{\rm end}$, $T_{\rm end}$, is 8.56 Gyr. 
In the following, we mainly use the age of galaxies to describe 
their evolution.  
Throughout the evolution of disk galaxies, the star formation rate 
is assumed to be proportional to gas mass fraction ($f_g$) of galaxies; 
\begin{equation}
\psi(t)=kf_g,
\end{equation}
where $k$ is a parameter which controls the star formation rate. 
This parameter $k$ is given as follows: 
\begin{equation}
k = \left\{
\begin{array}{lcl}
k_{\rm disk} \; \; \; & & {\rm for} \; \; 0 \le T < T_{\rm sb}, \\
k_{\rm sb}   & & {\rm for} \; \; T_{\rm sb} \le T < T_{\rm end}, \\
0            & & {\rm for} \; \; T_{\rm end} \le T.
\end{array}
\right.
\end{equation}
In the following, we refer the first, second, and third  phases 
as the prestarburst, 
the starburst, and the poststarburst, respectively. 
In the present study, we set the value of $k_{\rm disk}$ as 0.225 $\rm Gyr^{-1}$, 
which corresponds to a plausible star formation rate 
for Sb disks (e.g., Arimoto, Yoshii, \& Takahara 1992). 
$k_{\rm sb}$ is considered to be 10 times larger than $k_{\rm disk}$ 
($2.25 \; {\rm Gyr}^{-1}$), 
which is consistent with observational results on starburst 
galaxies (e.g., Planesas et al. 1997). 
Figure~1 shows the star formation history of our model. 

\placefigure{fig-1}

We try to understand the role of the 
{\it selective dust extinction} in the formation of e(a) galaxies by 
comparing the following two models with each other; 
a model with no dust extinction (from now on referred to as the $ND$ model) 
and the selective dust extinction model (the $SD$ model). 
In the $ND$ model, dust effects on photometric and spectroscopic 
properties are completely neglected; i.e., $A_V=0$ at all times. 
In the $SD$ model, we assume that during starburst 
the value of $A_V$ depends 
on the age of stellar population; i.e., 
\begin{equation}
A_V=A_{V,0} \exp \{ (T-t)/\tau \} \; \; \; 
{\rm for} \; \; T_{\rm sb} < T < T_{\rm end}, 
\end{equation}
where $A_{V,0}$ and $\tau$ are parameters controlling the 
degree of extinction. 
We here  set $A_{V,0}=5$ mag and $\tau=1.0 \times 10^6$ yr. 
 
Based on the monochromatic flux derived in the equation (1) 
and the value of $A_V$ from the equation (4)  
for each age of a galaxy $T$, 
we calculated the SED of a galaxy corrected by dust extinction 
using the extinction law derived 
by Cardelli et al. (1989) 
and adopting the so-called screen model. 
For deriving the fluxes for various gaseous emission lines 
(H$\delta$ and [OII]) in dusty starburst galaxies, we first 
calculate the number of Lyman continuum photons, $N_{\rm Ly}$, 
by using the SED that are not modified by dust extinction. 
If all the Lyman continuum photons are used for ionizing the surrounding gas, 
the luminosity of H$\beta$ is calculated according to  the following formula; 
\begin{equation}
L({\rm H}\beta) ({\rm erg \; s^{-1}}) = 4.76 \times 10^{-13} N_{\rm Ly} ({\rm s^{-1}})
\end{equation}
(Leitherer \& Heckman 1995). 
To calculate luminosities of other emission lines, 
e.g., [OII] and H$\delta$, we use the relative luminosity 
to H$\beta$ luminosity tabulated in PEGASE 
(Fioc \& Rocca-Volmerange 1997) which is calculated 
for the set of electron temperature of 10000 K and 
electron density of 1 cm$^{-3}$. 
Thus the SED derived in the present study consists of 
stellar continuum and gaseous emission. 


\placefigure{fig-2}
\placefigure{fig-3}

\section{Results}

Figure~2 shows the time evolution of EW(H$\delta$) and EW([OII]) 
for the $ND$ model and the $SD$ one. 
First, we describe the evolution for the $ND$ model. 
The equivalent width of [OII] decreases gradually 
before the starburst begins (i.e., at $T < 7.64 \; {\rm  Gyr}$). 
At $T=T_{\rm sb}$, the equivalent width of [OII] increases rapidly 
following the rapid increase of star formation rate due to starburst 
and then decreases gradually with the decreasing star formation rate.  
During the starburst, the equivalent width of [OII] is larger than 40 \AA. 
After star formation ceases at $T=T_{\rm end}$ and  massive stars 
formed during starburst die out, 
the equivalent width of [OII] rapidly 
decreases and becomes smaller than 5 \AA. 
The strength of equivalent width of H$\delta$ evolves as follows. 
During the prestarburst evolution phase, 
the H$\delta$ line is observed as an absorption one and 
its equivalent width reaches  the local maxima at around 3 Gyr 
($z \sim 1.5$). 
The strength of EW(H$\delta$) gradually decreases, 
and  becomes  about 3 \AA~ just before
starburst begins ($T = 7.64 \; {\rm Gyr}$, $z = 0.4$). 
When starburst begins, the H$\delta$ line changes from an absorption line 
to an emission one 
and its equivalent width reaches $-5$ \AA.  
${\rm A \; few} \times 10^8 \; {\rm  yr}$ after the starburst begins, 
H$\delta$ line changes into absorption line  again. 
Since the flux of H$\delta$ emission line 
becomes negligibly small, the equivalent width of H$\delta$ reaches 8 \AA~
when starburst ceases.
As a result of passive evolution of galaxies after starburst, 
the equivalent width of H$\delta$ decreases gradually: 
it becomes smaller than 7 \AA~ at 8.8 Gyr and 3 \AA~ at 11 Gyr. 

Next we describe the evolution of the $SD$ model. 
Since we assume that the selective extinction affects spectroscopic
properties of galaxies  only in  
starburst phase, 
there is no difference in the evolution
of EW([OII]) and EW(H$\delta$) between the $SD$ model and the $ND$ one 
in the prestarburst phase and the poststarburst one. 
We therefore pay our attention to  the evolution 
of EW([OII])  and EW(H$\delta$)  during starburst. 
In the $SD$ model, the equivalent width of [OII] does not increase 
(but increases  in the $ND$ model) 
and is below 40 \AA~ 
during  starburst. 
The equivalent width of H$\delta$ becomes rapidly small  
just after starburst begins, 
although the H$\delta$ line can be  still observed as an absorption line. 
A few $\times 10^8$ years after the starburst begins, 
the equivalent width of H$\delta$ becomes larger than 4 \AA. 
The reason why the equivalent width of H$\delta$ of the $SD$ model 
is always larger than that of the $ND$ model is that 
the flux of H$\delta$ emission line of the  $SD$ model is smaller 
than that of $ND$ model. 

Figure~3 shows the evolution of galaxies on EW([OII])-EW(H$\delta$) 
plane for each model. 
The criteria of spectroscopic classification determined by 
Dressler et al. (1999) are also superimposed  on it. 
The $ND$ model  is located  within the  e(b) region during starburst   
and is very rapidly shifted to  the a+k one  just when starburst is ended. 
Although, the locus of the $ND$ model passes through the e(a) region, 
the crossing time is $4 \times 10^6$ years which is about 0.5 \% 
of the duration of the starburst. 
On the other hand, the $SD$ model can successfully show the spectroscopic 
properties of e(a) galaxies, namely, strong H$\delta$ absorption line 
[EW (H$\delta$) $> 4$ \AA] 
and relatively modest [OII] emission line [EW([OII]) $>$ 5 \AA], 
during the most of the starburst phase (0.7 Gyr). 
This result accordingly suggests that a selective extinction plays a vital 
role in the formation of galaxies with e(a) spectra. 
These results thus  confirm the early suggestion by Poggianti \& Wu 
(2000) that if the youngest stellar population is the most heavily obscured 
by dust in a dusty starburst galaxy, the e(a) spectra 
can be achieved in the galaxy.

As is described above,
we have confirmed that the  $SD$ model shows 
the weaker [OII] emission and stronger H$\delta$ absorption 
compared  with the $ND$ model. 
The physical reason for the successful reproduction
of the $SD$ model can be  understood in terms of 
the difference in the influence of dust extinction  
between continuum and emission lines. 
The emission lines come from the ionized gas  
around youngest stellar populations which are most heavily 
obscured by dust. 
On the other hand, most of the continuum
come from the stellar populations  
that are older and less obscured than younger ones  dominating  the 
ionizing photons.   
The flux of emission lines consequently is more greatly
affected by dust extinction  than 
that of stellar continuum. 
Thus the $SD$ model can 
show the smaller equivalent width of emission lines 
without changing 
so greatly the flux of continuum.


\section{Discussion and conclusions}

Poggianti \& Wu (2000) first discussed that if the dust extinction for 
a stellar population decreases with the stellar age, e(a) spectra 
can be obtained for a starburst galaxy. 
They furthermore suggested that if the location and thickness of dust 
patches depend on the age of the embedded stellar populations in a 
starburst galaxy, the effects of the above selective dust extinction become 
remarkable for the galaxy. 
Then, when and how the proposed difference in the degree of dust 
extinction between stellar populations with different ages is possible 
during galactic evolution in distant clusters?
We here suggest that the difference of dust extinction between the central 
region and the outer one in a galaxy is
 a main cause for the above age-dependent 
extinction. 
To be more specific, since younger stellar populations formed by 
secondary starburst in the central region with a larger amount 
of dusty interstellar gas are more heavily obscured by dust than the 
outer old components, the age-dependent extinction 
can be achieved. 
This radial dependence of dust extinction is considerably reasonable and 
realistic, considering that secondary dusty starburst occurs preferentially 
in the central part of a galaxy owing to efficient inward gas 
transfer driven by non-axisymmetric structure (such as stellar bars) 
and galaxy interaction and merging. 
Numerical simulations on gaseous and stellar distribution in merging 
disk galaxies with dusty starburst furthermore have demonstrated that 
the central young stellar component formed by nuclear starburst is more 
heavily obscured by dusty gas than the outer old component initially 
located in merger progenitor disks (Bekki, Shioya, \& Tanaka 1999). 
Thus we strongly suggest that galaxies with the central dusty starburst 
are more likely to show the selective dust extinction. 
Future observational studies on the radial dependence of photometric and 
spectroscopic properties for e(a) galaxies, which can reveal the 
detailed distribution of young dusty population and that of old one 
with less extinction, will assess the validity of the above idea 
for the origin of the selective dust extinction (i.e., the age-dependent extinction). 

Then what physical process is closely associated with the radial 
dependence of dust extinction suggested above and thus with the 
formation of e(a) galaxies in distant clusters ? 
We here suggest that minor and unequal-mass galaxy merging, which 
is different from major merging in that it can leave disk systems 
even after merging, is {\it one} of possible candidates that can 
reproduce fundamental properties of e(a) galaxies. 
Mihos \& Hernquist (1994) has demonstrated that minor merging can 
excite non-axisymmetric structure in a late-type disk, trigger 
nuclear starburst, and transform the disk into an early-type disk such 
as Sa and S0. 
Unequal-mass merging between two disks is furthermore demonstrated to 
create a disk with the structure and kinematics strikingly similar to those 
of S0s (Barnes 1994), and furthermore Bekki (1998) found that 
such merging can trigger nuclear starburst, form a very gas-poor disk, 
and create a remarkable bulge characteristic of typical S0s. 
These merging therefore not only contribute to the formation of 
dusty nuclear starburst and thus to the radial dependence of dust 
extinction but also provide an evolutionary link between late-type 
disks, e(a) galaxies, and early-type disks (S0) with the number fraction 
observed to decrease with redshift (e.g., Dressler et al. 1996). 
An early-type disk galaxy formed by minor and unequal-mass merging 
experiences the post-starburst phase with k+a/a+k spectra after starburst 
and shows a remarkable bulge with young stellar populations and 
thus with bluer colors (Bekki 1998). 
Accordingly one of observational tests for assessing
the relative importance of 
minor and unequal-mass merging in the fraction of e(a) galaxies with 
the selective dust extinction is to investigate the number fraction
of  disk galaxies 
with k+a/a+k spectra and prominent blue bulges 
in distant clusters.

\acknowledgments
We are  grateful to the anonymous referee for valuable comments,
which contribute to improve the present paper.
Y.S. thanks to the Japan Society for Promotion of Science (JSPS) 
Research Fellowships for Young Scientist.

\clearpage

\figcaption{
The star formation history of our model. 
Starburst begins at $T_{sb}=7.64$ Gyr  ($z_{\rm sb}$ = 0.4) and
stops at $T_{end}=8.56$ Gyr ($z_{\rm end}$ = 0.29). 
We here assume that the mass of the model galaxy is 
$6 \times 10^{10}M_{\odot}$ (corresponding to the
disk mass of the Galaxy).
\label{fig-1}}

\figcaption{
Upper panel: The evolution of the equivalent width of [OII] 
emission line [EW([OII])]. 
The $ND$ model (the model without dust extinction) and the $SD$ one
(with selective dust extinction)
 are shown as dotted line and solid line, respectively. 
Lower panel: The evolution of the equivalent width of H$\delta$ 
[EW(H$\delta$)]. 
Red lines, blue lines, and black lines mean 
the equivalent width of the stellar absorption, 
the equivalent width of emission lines form ionized gas, and 
the sum of both lines, respectively. 
Positive (negative) EW(H$\delta$) in black lines 
means that H$\delta$ line 
is observed as absorption (emission) line. 
\label{fig-2}}

\figcaption{
The evolution of galaxies on the EW([OII]) - EW(H$\delta$) plane. 
The $ND$ model and the $SD$ one 
are shown as a dotted line and solid one, respectively. 
Some points (T=1, 7.64, 8.56, 8.58, 12 Gyr) are shown 
as open circles (the $ND$ model) and filled circles (the $SD$ model) 
with age. 
The criteria of classification by Dressler et al. (1999) 
are also superimposed.
Note that only the $SD$ model can pass through the e(a) region 
during starburst (7.64 Gyr $< T <$ 8.56 Gyr),
which implies that the selective dust extinction during starburst
is very important for the formation of e(a) galaxies. 
\label{fig-3}}


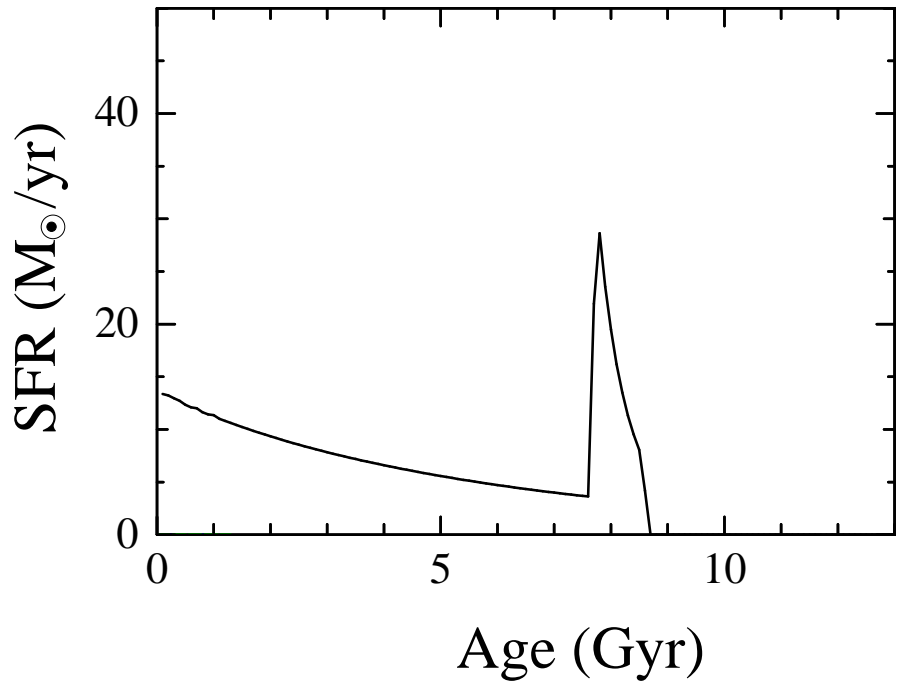
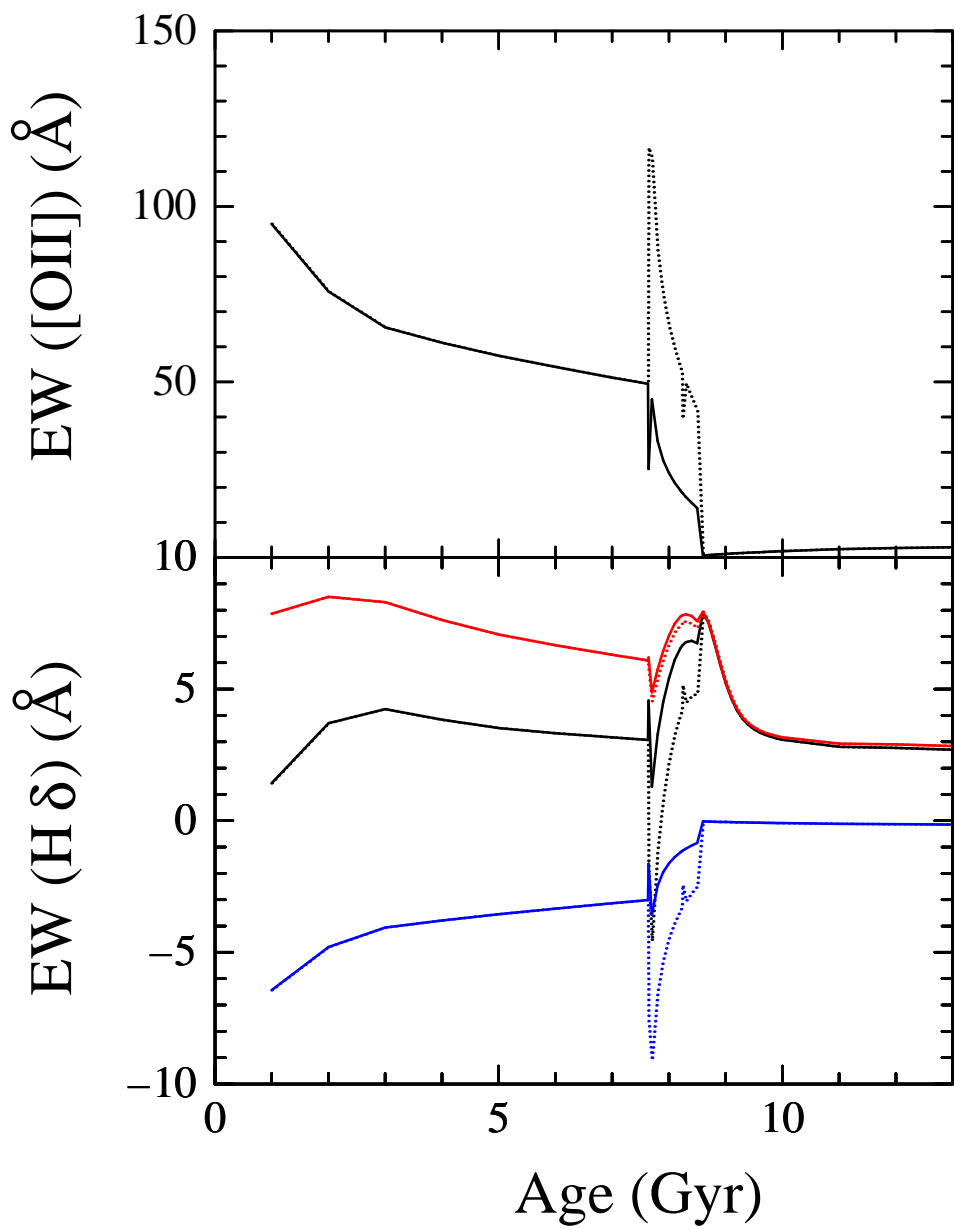
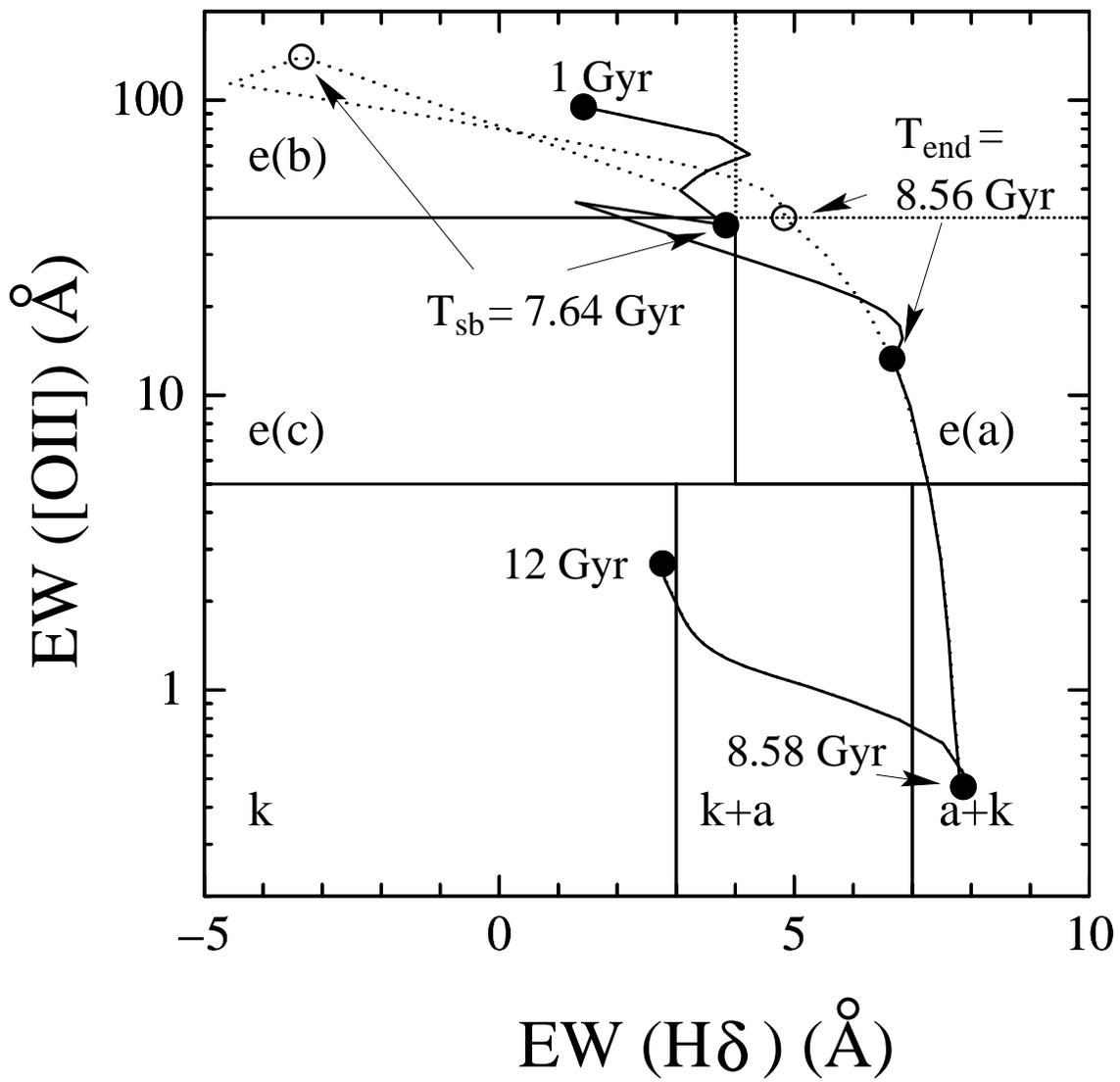
\end{document}